\documentclass[aps,manuscript,showpacs,showkeys,superscriptaddress,nofootinbib]{revtex4-1}
\usepackage{graphicx}
\usepackage{epsfig}  
\usepackage{epsf}    
\usepackage{dcolumn}
\usepackage{bm}
\usepackage{dcolumn}
\usepackage{textcomp}
\usepackage[tbtags]{amsmath}
\usepackage{amsfonts}
\usepackage{float}
\usepackage{subfig}

\usepackage[]{hyperref}
  \hypersetup{
  bookmarks=true,         
  unicode=true,         
  pdftoolbar=true,     
  pdfmenubar=true,     
  pdffitwindow=true,    
  pdfstartview={FitH},    
  pdfsubject={Sterilenus@DUNE},  
  pdfnewwindow=true,     
  pdfcreator={RevTeX},
  colorlinks=true,     
  linkcolor=red,   
  citecolor=blue,    
  filecolor=black,  
  urlcolor=blue,           
  }
\usepackage{hypcap}

\newcommand{\beq}{\begin{equation}}
\newcommand{\eeq}{\end{equation}}
\newcommand{\beqa}{\begin{eqnarray}}
\newcommand{\eeqa}{\end{eqnarray}}

\newcommand{\ty}{{\theta_{13}}}
\newcommand{\tz}{{\theta_{23}}}

\newcommand{\dcp}{\delta_{\mathrm{CP}}}

\begin{document}
\title{Effect of Non Unitarity on Neutrino Mass Hierarchy determination at DUNE, NO$\nu$A and T2K }
\author{Debajyoti Dutta}
\email[Email Address: ]{debajyotidutta@hri.res.in}
\affiliation{Harish-Chandra Research Institute, Chhatnag Road, Jhunsi, Allahabad 211019, India}
\affiliation{Homi Bhabha National Institute, Training School Complex,
Anushaktinagar, Mumbai - 400094, India}

\author{Pomita Ghoshal}
\email[Email Address: ]{pomita.ghoshal@gmail.com }
\affiliation{Department of Physics, LNM Institute of Information Technology (LNMIIT),\\
Rupa-ki-Nangal, post-Sumel, via-Jamdoli, Jaipur-302 031, Rajasthan, India}

\author{Samiran Roy$^{1, 2,}$}
\email[Email Address: ]{samiranroy@hri.res.in}
\date{\today}

\begin{abstract}
  { The neutrino mass ordering is one of the principal unknowns in the neutrino sector. Long baseline neutrino experiments have the potential of resolving this issue as they are sensitive to large matter effects. The superbeam experiment DUNE is one of the most promising candidates to study the neutrino mass hierarchy, along with NO$\nu$A and T2K. But in the presence of non unitarity of the leptonic mixing matrix, the capability of such experiments to discriminate between the two hierarchies  gets suppressed. The mass hierarchy sensitivity of DUNE decreases in the presence of new physics. In this paper we analyze the origin and extent of this loss of sensitivity at the level of oscillation probabilities, events, mass hierarchy sensitivity and the discovery reach of DUNE, NO$\nu$A and T2K.} 
\end{abstract}


\maketitle


\section{Introduction}
\label{Introduction}
  In the three flavor neutrino oscillation scenario, the oscillations are driven by the two mass-squared differences $\Delta m^2_{21}$ and $\Delta m^2_{31}$ and the oscillation amplitudes are measured by the three mixing angles $\theta_{12}$, $\theta_{13}$ and $\theta_{23}$. 
The leptonic CP phase $\delta_{cp}$ is still the least known parameter in this scheme. 
The CP violation in the leptonic sector may lead to leptogenesis \cite{joshipura, endoh, pascoli, Branco, pascoli2} and can explain the observed matter-antimatter asymmetry in the Universe. 
Further, it is not yet confirmed whether the atmospheric mixing angle $\theta_{23}$ lies in the higher octant (HO) i.e. $\theta_{23}$ $ >$ 45$^0$ or in the lower octant (LO) i.e.  $\theta_{23}$ $ <$ 45$^0$. Present results from the T2K experiment still prefer maximal $\theta_{23}$ \cite{t2k_maximal}, while the results from the NO$\nu$A collaboration exclude the maximal mixing scenario at 2.5$\sigma$ C.L. \cite{nova1}. So the octant of $\theta_{23}$ and its deviation from maximal mixing is still to be established with a high degree of precision.

 The ordering of the neutrino masses is another critical unknown in this scheme. In long baseline neutrino experiments, the earth matter effect plays an important role. The matter effect has opposite signs for the two hierarchies in the probability expression . So experiments like DUNE \cite{dune1,dune2}, LBNO \cite{lbno}, HK \cite{hk}, NO$\nu$A\cite{nova11}, T2K \cite{t2k1, t2kNew} etc have the potential of distinguishing between the normal (NH, $m^2_{3}-m^2_{1}>0$) and inverted (IH, $m^2_{3}-m^2_{1} < 0$) mass hierarchies. Since the neutrino mass models developed for the two hierarchies are significantly different, a prior knowledge of the mass hierarchy may help in discriminating between classes of models and hence can assist in formulating a more definitive picture of neutrino masses. The neutrino mass hierarchy has been studied in different context in \cite{gandhi1, gandhi2,gandhi3,sandhya, prakash, masud, dev123, subha1, subha2, raut1, raut11, Zhang, ohs, x}. 
  
There is an additional complication which may hamper the determination of the neutrino mass ordering - the possible presence of new physics which can give rise to additional CP phases. The new phases may mimic the leptonic CP phase and lead to further degeneracies. Non-unitarity (NU) in the neutrino mixing matrix is one of the possible departures from the standard three-neutrino mixing framework. One of the possibility of NU is due to the induction of neutrino mass through the type-I seesaw mechanism. If the messenger fermions involved are within the reach of the Large Hadron Collider, a rectangular leptonic mixing matrix would be obtained, giving an effectively non-unitary neutrino mixing matrix \cite{ antu, fern, Goswami, Escrihuela, Antusch}. This framework has a new non-unitary phase which is degenerate with the standard CP phase and affects the sensitivity to CP violation \cite{me}. In \cite{Miranda}, this degeneracy was discussed at the level of oscillation probabilities, and in \cite{Pasquini} a solution was suggested through the upgrade of T2HK to TNT2K. The effect of non unitarity was also studied at the probability level in vacuum for the T2K, NOVA and DUNE experiments in \cite{gmiranda}. In \cite{qian}, the unitarity of the PMNS matrix was tested using direct and indirect methods. Recent constraints related to source- detector NSI can be found in \cite{amir}.

 The effect of non-unitarity on neutrino mass hierarchy measurements has not been extensively studied yet. In this work, we explore the effect of non-unitarity on measurements of the neutrino mass ordering. This study has been performed in the context of the three long baseline (LBL) neutrino experiments T2K, NO$\nu$A and DUNE.  We analyze the effect of non-unitary mixing on the oscillation probabilities for the given experiment baselines, and describe the mass hierarchy sensitivity for the individual experiments. 
We show that the hierarchy sensitivity decreases in the presence of non unitarity. The experiments are simulated using the standard long baseline package GLoBES \cite{phuber1, phuber2},
which includes earth matter effects and relevant systematics for each experiment. We have used MonteCUBES's \cite{Blennow} Non Unitarity Engine (NUE) with GLoBES while performing this analysis.


The paper is organized as follows: in Section II we discuss the oscillation probability $P (\nu_{\mu} \rightarrow \nu_{e})$ relevant to the given experiments in the presence of non-unitary mixing. Section III gives some information regarding the experiments NO$\nu$A, T2K and DUNE, and outlines the simulation procedure followed by us to compute the mass hierarchy sensitivity for the experiments. In Section IV, we give bi-probability and bi-event plots for the relevant baselines. In section V, we present the results for the mass hierarchy sensitivity as well as the mass hierarchy discovery potential of these experiments in the presence of non unitarity. In Section VI the results are discussed and conclusions are drawn. 
 
\section{Effect of a Neutral Heavy Lepton in Neutrino Oscillations}

The model of non unitarity used in this work is based on \cite{valle1987}. The  symmetrical parametrization technique can be found in \cite{valle1980}. In the presence of a Neutral Heavy Lepton, the $3\times 3$ neutrino mixing matrix does not remain unitary and instead becomes  \begin{equation}N = N^{NP}U\end{equation}, where U is the $3\times 3$ PMNS matrix.  The left triangular matrix $N^{NP}$ can be written as \cite{Escrihuela}
 
 \begin{equation}
N^{NP} = 
\begin{pmatrix}
\alpha_{11} & 0 & 0 \\
\alpha_{21} & \alpha_{22} & 0 \\
\alpha_{31} & \alpha_{32} & \alpha_{33} \\
\end{pmatrix}
\end{equation}

 In the presence of non unitarity matrix, the electron neutrino appearance probability changes in vacuum, as explained in \cite{Escrihuela, Pasquini}. The expression for $P_{\mu e}$ with NU can be written as
\begin{equation}
P_{\mu e} = (\alpha_{11}\alpha_{22})^2 P^{3\times 3}_{\mu e}+\alpha_{11}^2\alpha_{22}|\alpha_{21}|P^{I}_{\mu e}+\alpha_{11}^2|\alpha_{21}|^2
\end{equation}, where $P^{3\times 3}_{\mu e}$ is the standard three flavor neutrino oscillation probability and $P^{I}_{\mu e}$ is the oscillation probability containing the extra phase due to non unitarity in the mixing matrix. $P^{3\times 3}_{\mu e}$ above can be written as :
 \begin{equation}
 \begin{split}
 P^{3\times 3}_{\mu e} = 4[\cos^2\theta_{12}\:\cos^2\theta_{23}\:\sin^2\theta_{12}\:\sin^2(\frac{\bigtriangleup m^2_{21}L}{4E_{\nu}})+\cos^2\theta_{13}\:\sin^2\theta_{13}\:\sin^2\theta_{23}\:\sin^2(\frac{\bigtriangleup m^2_{31}L}{4E_{\nu}})] \\ +\sin(2\theta_{12})\:\sin\theta_{13}\:\sin(2\theta_{23})\:\sin(\frac{\bigtriangleup m^2_{21}L}{2E_{\nu}})\:\sin(\frac{\bigtriangleup m^2_{31}L}{4E_{\nu}})\:\cos(\frac{\bigtriangleup m^2_{31}L}{4E_{\nu}}-I_{123})
 \end{split}
  \end{equation}
 
 And
  \begin{equation}
 \begin{split}
 P^{I}_{\mu e} = -2[\sin(2\theta_{13})\:\sin\theta_{23}\:\sin(\frac{\bigtriangleup m^2_{31}L}{4E_{\nu}})\:\sin(\frac{\bigtriangleup m^2_{31}L}{4E_{\nu}}+\phi_{21}-I_{123})]\\
 -\cos\theta_{13}\:\cos\theta_{23}\:\sin(2\theta_{12})\:\sin(\frac{\bigtriangleup m^2_{21}L}{2E_{\nu}})\sin({\phi_{21}})
  \end{split}
  \end{equation}
  where
 $I_{123} = -\delta_{cp}$ and $\alpha_{21} = |\alpha_{21}|\exp(\phi_{21})$. Here we have observed that only four extra parameters from $N^{NP}$ enter the vacuum probability expression for $P_{\mu e}$- the real parameters $\alpha_{11}$ and $\alpha_{22}$, one complex parameter $|\alpha_{21}|$ and the phase associated with $|\alpha_{21}|$. In our analysis, we have not considered the effect of the third row elements of $\rm N^{NP}$ matrix as their contributions are negligible even in the presence of matter effect.
\section{Simulation Parameters and Experiment details}

In this work, we have studied the neutrino mass hierarchy sensitivity of three long baseline experiments- T2K (Tokai to Kamioka), NO$\nu$A (The NuMI{\footnote{Neutrinos at the Main Injector} } Off-axis $\nu_{e}$ Appearance experiment) and DUNE (Deep Underground Neutrino Experiment). The main goal of T2K is to observe
$\nu_{\mu}\rightarrow\nu_{e}$ oscillations and to measure $\ty$ as well as leptonic CP violation (T2K new data gives hint of leptonic CP violation \cite{t2k_cp}) while NO$\nu$A can measure the octant of $\tz$, the neutrino mass hierarchy, $\ty$ and leptonic CP violation. DUNE, with its 1300 km baseline, can address all these issues with a higher degree of precision. In our recent work \cite{me}, we have studied the CP violation sensitivity and discovery reach of these experiments in the presence of non unitarity. We have specified all the experimental and simulation details in that work, and will be using the same informations for this work also.

Here, we fix the three-flavor neutrino oscillation parameters to their best fit values taken from \cite{Gonzalez-Garcia}. Since the solar and reactor mixing angles are the most precisely measured, we take $\theta_{12} = 33.48^0$ and $\theta_{13} = 8.5^0$ \cite{Gonzalez-Garcia} respectively. For true NH (IH), the value of the two mass square differences are $\Delta{m}^2_{21} = 7.5 \times 10^{-5}$ e$V^2$ and $\Delta{m}^2_{31} = 2.457 \times 10^{-3}$ e$V^2$ ($-2.449 \times 10^{-3}$ e$V^2$) respectively. No priors are added on any of the parameters. Again, the octant issue is not yet resolved and different global analyses prefer different octant \cite{Gonzalez-Garcia,Capozzi, Forero1} as the true octant. In this work, we consider the maximal value of $\tz$ as the true value i.e. $\theta_{23} = 45^0$. However, the physics conclusions drawn in this work are not going to change significantly even if we consider non maximal $\theta_{23}$ in `data' and then marginalize it in `fit' in the allowed 3$\sigma$ range. The effect seen in the probability level at DUNE [See Appendix 1] can be realised at more than 5$\sigma$ CL only. This point is discussed in the text.

In the literature, there are a few studies available regarding the constraints on non unitarity parameters \cite{Escrihuela, Antusch, enrique1, enrique2}. Universality constraints give strong bounds on the diagonal NU parameters which in turn give further restrictions on the off diagonal parameters \cite{enrique1, enrique2}. Since these bounds are derived considering the charged current induced processes with the assumption that there is no new physics other than the non-unitarity mixing coming from type I see-saw, hence in the presence of other new physics e.g. right-handed interactions or neutrino-scalar Yukawa interactions in type II see-saws, these bounds are not valid. On the other-hand, the off diagonal NU parameters are directly restricted by the neutrino experiments like NOMAD \cite{nomad} and CHORUS \cite{chorus1, chorus2}, and these bounds are less stringent and model-independent. In this work, we do not adopt the approach which would use the stronger, but model-dependent bounds given in \cite{enrique1, enrique2}, but instead, use the model-independent bounds from neutrino experiments \cite{Escrihuela}. The bounds that we use in this work are: $\alpha_{11}^2 \geq 0.989$, $\alpha_{22}^2 \geq 0.999$ and $|\alpha_{21}|^2 \leq 0.0007$ at 90\% C.L. \cite{Escrihuela}. The allowed range of $\phi_{21}$ is $[-\pi, \pi]$. We assume the limiting values of these NU parameters while generating the bi-probability and bi-event plots. In our $\chi^2$ analysis, the central values of the allowed ranges are taken as the true values (unless stated). 

\section{Bi-probability and Bi-event plots}

\begin{figure}[h] 
\centering
\includegraphics[width=0.48\textwidth]{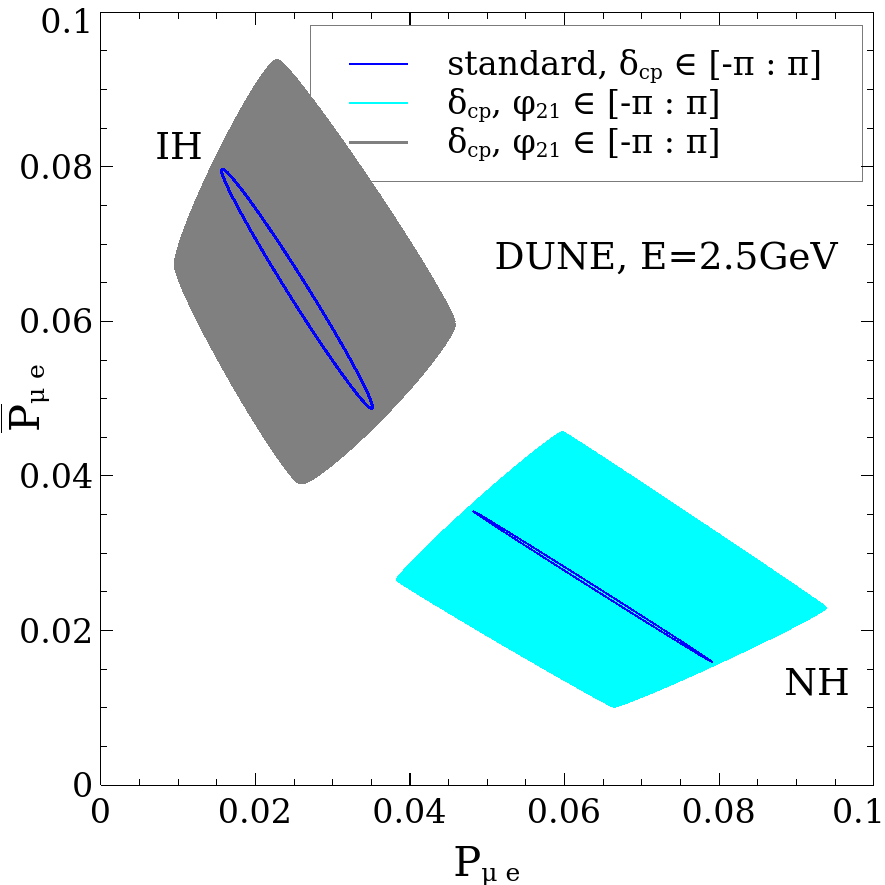}
\includegraphics[width=0.48\textwidth]{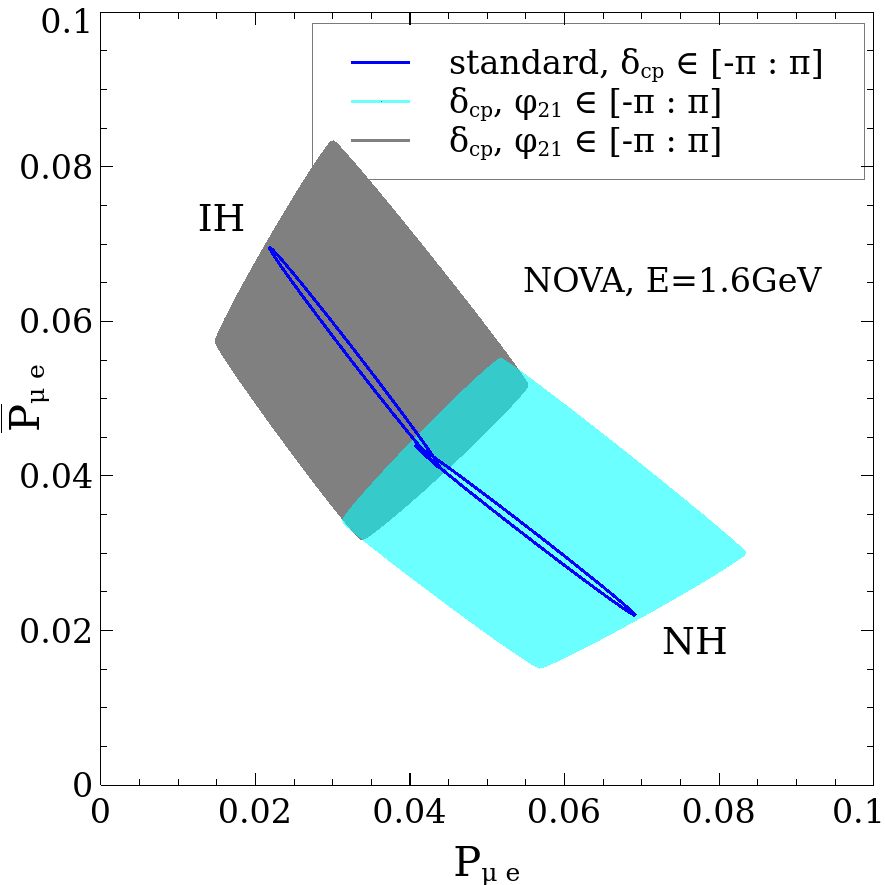}
\includegraphics[width=0.48\textwidth]{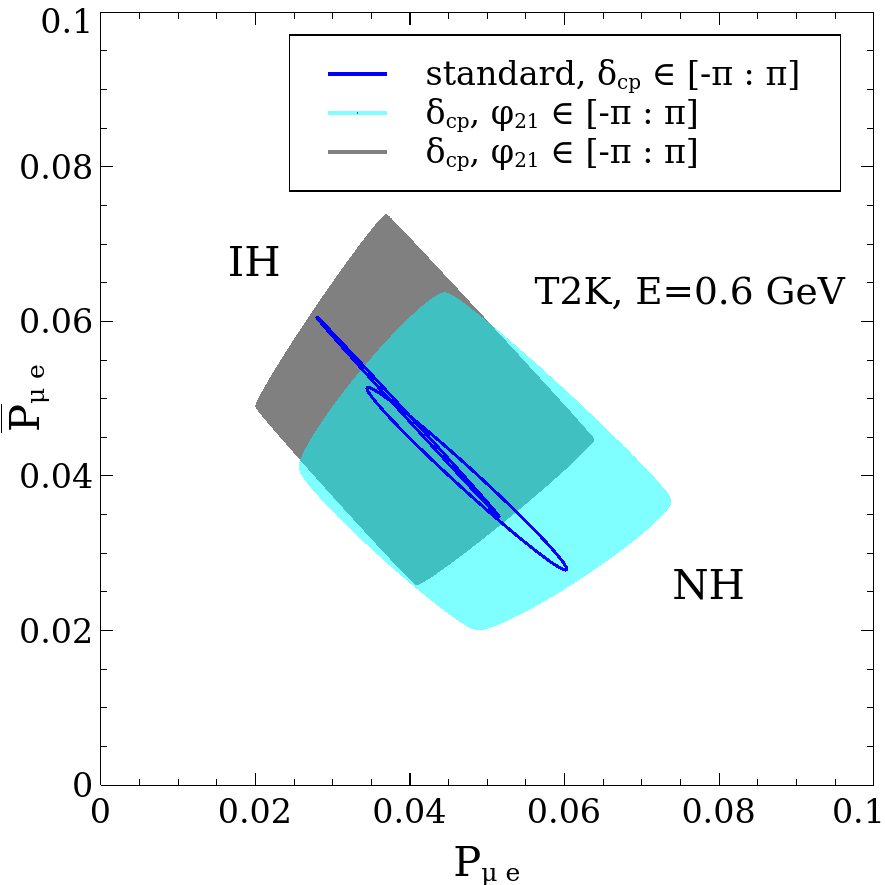}
\caption{\footnotesize{ Bi-probability plots ($\bar{P}$ versus $P$): For DUNE, NO$\nu$A and T2k at their peak energy. The blue ellipse corresponds to the standard 3$\nu$ case and is obtained by varying $\dcp \in [-\pi, \pi]$. The cyan and the gray bands show the effect of non unitarity for NH and IH when both the phase $\dcp$ and $\phi_{21}$ is varied from $-\pi$ to $\pi$ and all other NU parameters are fixed at their limiting values i.e. $\alpha_{11} = 0.9945$, $\alpha_{22} = 0.9995 $ and $|\alpha_{21}| = 0.0257$. The green and the red ellipses are two special cases in both the hierarchies. }}

\label{prob}
\end{figure}

In this section, we study the effect of non-unitarity at the probability level and explain the mass hierarchy degeneracy on the basis of bi-probability and bi-event plots. The bi-probability plots are shown in the $\rm P(\nu_{\mu}\rightarrow \nu_e )$ - $\rm P(\bar{\nu_{\mu}}\rightarrow \bar{\nu_e})$ plane. In figure \ref{prob}, the blue solid ellipse corresponds to the standard case with $\delta_{cp}$ varying from $-\pi$ to $\pi$. The gray and the cyan shaded regions (for Normal and Inverted Hierarchy respectively) correspond to the non-unitary case where the NU phase $\phi_{21}$ is also varied with $\delta_{cp}$ from $-\pi$ to $\pi$. The true values of the NU parameters (except $\phi_{21}$) are fixed at their upper and lower bounds. For each $\delta_{cp}$, due to the variation of $\phi_{21}$ in [$-\pi$, $\pi$], we get a continuous band of ellipses. From figure \ref{prob}, we draw the following conclusions:

\begin{itemize}

\item For DUNE at its peak energy i.e. $E = 2.5 GeV$, it is observed that in the 3$\nu$ framework, ellipses corresponding to NH and IH are well separated. Even in the presence of NU there is no overlapping between the ellipses that correspond to NH and IH. In the case of NO$\nu$A, there is some overlapping between the blue ellipses corresponding to NH and IH and the overlapping is more prominent in the presence of NU. In T2K, there is more overlap even for the 3$\nu$ case and with NU, the scenario worsens drastically. From this observation, we can conclude that only DUNE can discriminate between the two hierarchies at its peak energy even in the presence of NU.
\end{itemize}
The effect seen in figure \ref{prob} can also be verified from figure \ref{event1}. These bi-event plots are generated assuming the peak energy of the three experiments i.e. for DUNE, 
E = 2.5 GeV, for NO$\nu$A, E = 1.6 GeV and for T2K, E = 0.6 GeV. But to draw some realistic conclusions, we consider the bi-event plots between the energy integrated total number of neutrino and anti-neutrino events (figure \ref{event2}) for the three experiments. Here also, we fix the NU parameters to the boundary values and vary $\dcp$ and $\phi_{21}$ from $-\pi$ to $\pi$.

\begin{itemize}
\item  In DUNE, as in the bi-probability plots, the standard ellipses are well separated for NH and IH, indicating the ability of DUNE to resolve hierarchy degeneracy in the 3$\nu$ framework. But non unitarity induces a small degeneracy between the two hierarchies as there is a small overlapping between the gray and the cyan bands. If we consider the co-ordinate $(1000, 340)$ in the total events plot for DUNE, it lies in the overlapping region between the gray (IH) and the cyan (NH) band, and hence it is not possible to pinpoint the hierarchy near this co-ordinate. But due to the available spectral information as well as the high capability of the DUNE detector, DUNE may resolve the degeneracy seen in fig \ref{event2}. On the other hand, a co-ordinate say (2000, 300) in the cyan band is far removed from the standard blue ellipse as well as the overlapping region, and hence any event corresponding to this point can be a hint of NU with NH. The situation gets worse in the case of NO$\nu$A and T2K. The standard 3$\nu$ ellipses that correspond to NH and IH show a similar behavior as the corresponding bi-probability plots. The overlapping is more prominent in the presence of NU. If the leptonic mixing matrix is non unitary, then these experiments are unable to discriminate between the two hierarchies. From this point of view, non unitarity of the leptonic mixing matrix has to be taken seriously.
\begin{figure}[h]
\centering
\includegraphics[width=0.45\textwidth]{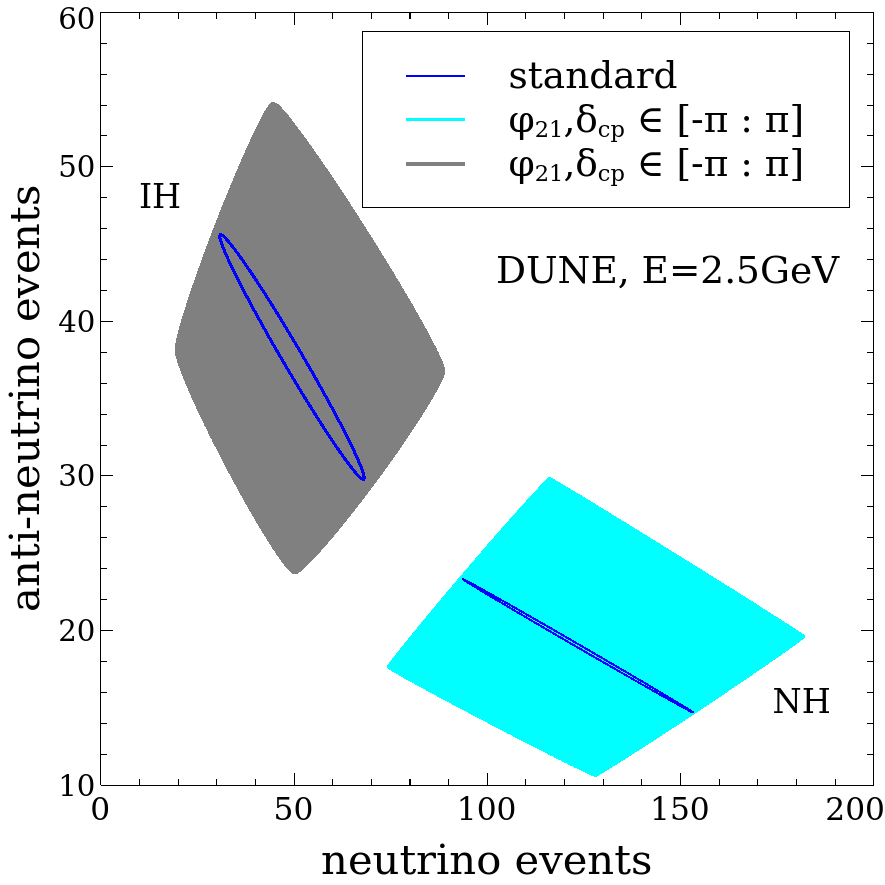}
\includegraphics[width=0.45\textwidth]{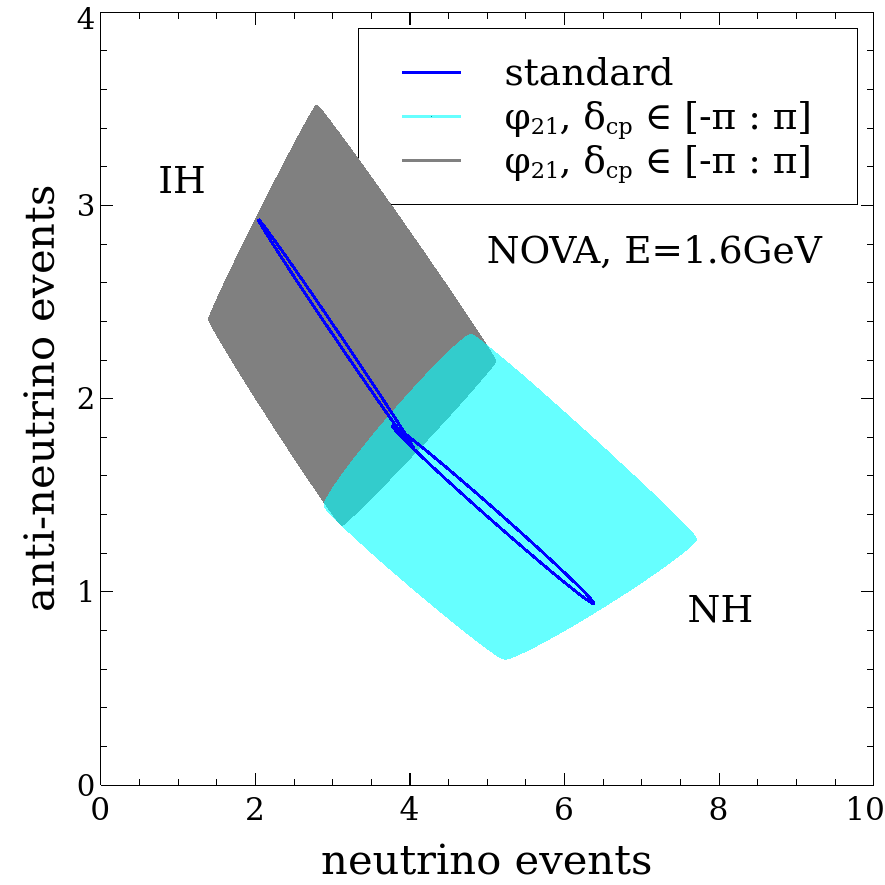}
\includegraphics[width=0.45\textwidth]{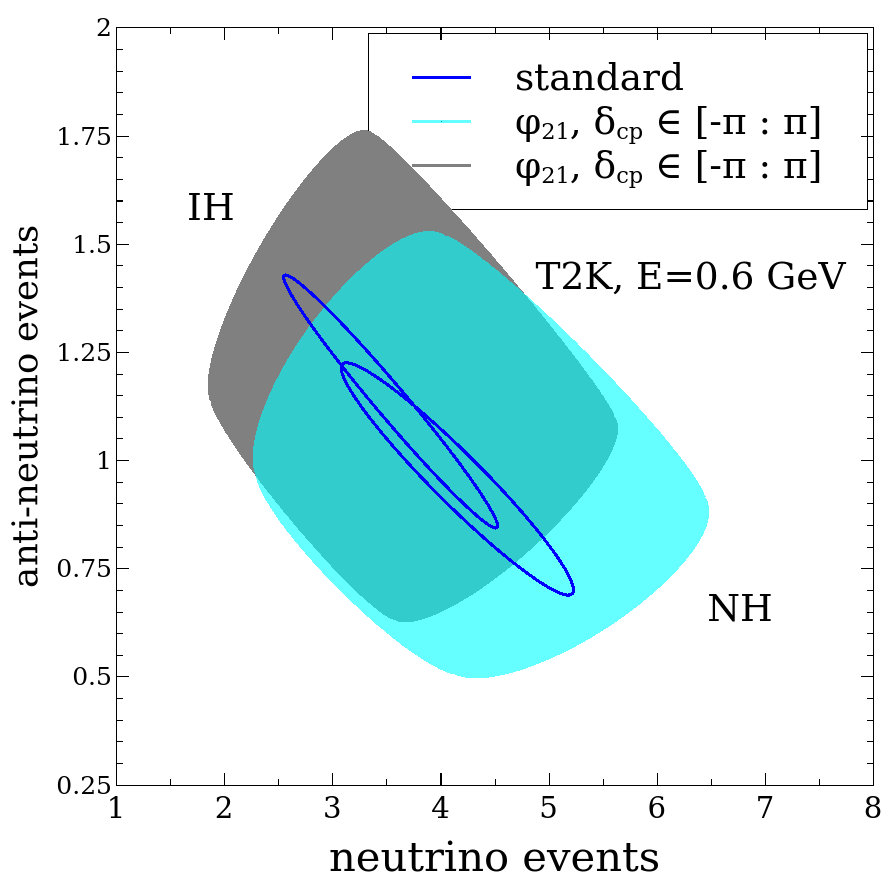}
\caption{\footnotesize{Bi-event plots For DUNE, NO$\nu$A and T2k at their peak energy. The blue ellipse corresponds to the standard 3$\nu$ case and is obtained by varying $\dcp \in [-\pi, \pi]$. The gray and the cyan band show the effect of non unitarity when both the phase $\dcp$ and $\phi_{21}$ is varied from $-\pi$ to $\pi$ and all other NU parameters are fixed to their limiting values.}}
\label{event1}
\end{figure}

\begin{figure}[h] 
\centering
\includegraphics[width=0.48\textwidth]{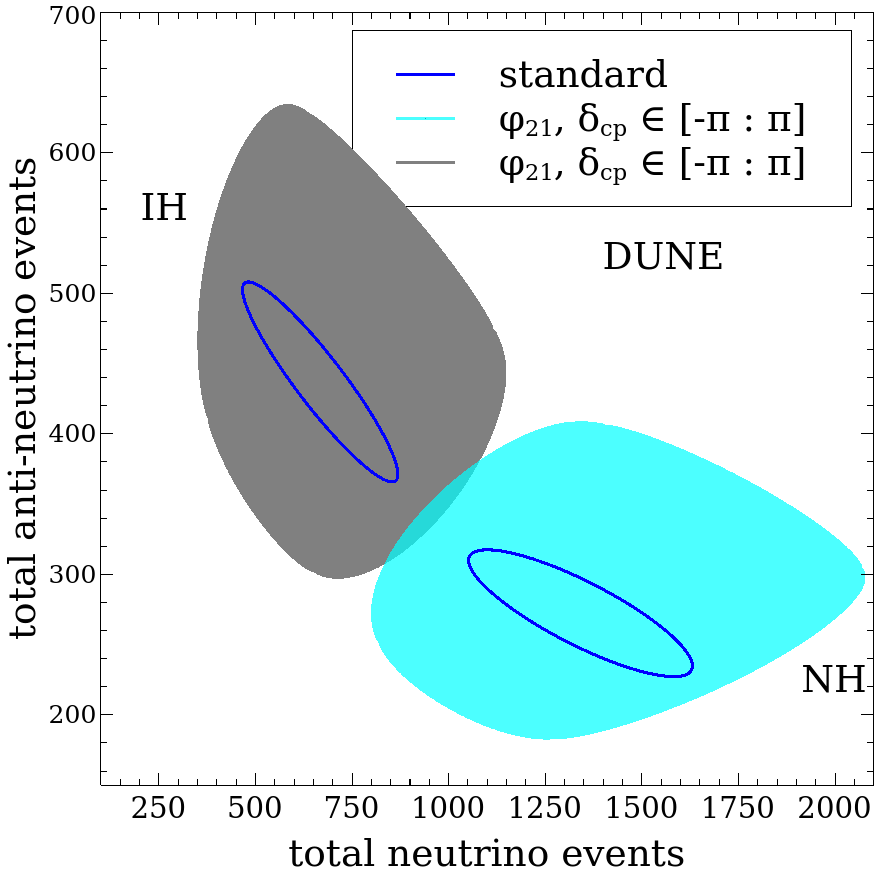}
\includegraphics[width=0.48\textwidth]{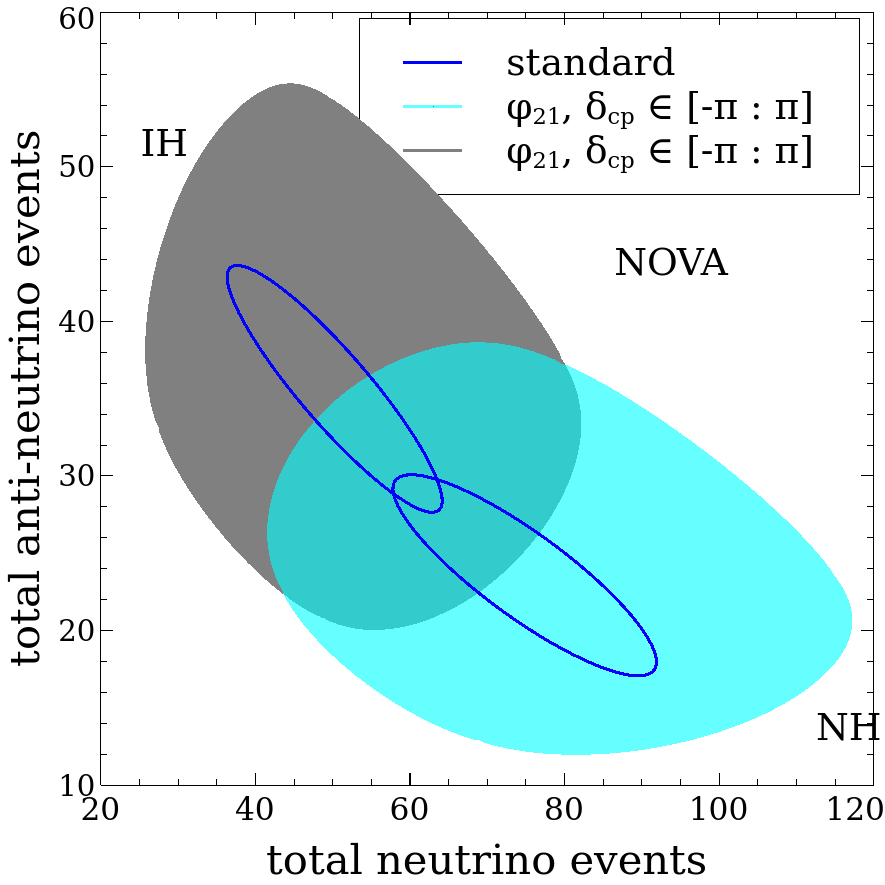}
\includegraphics[width=0.48\textwidth]{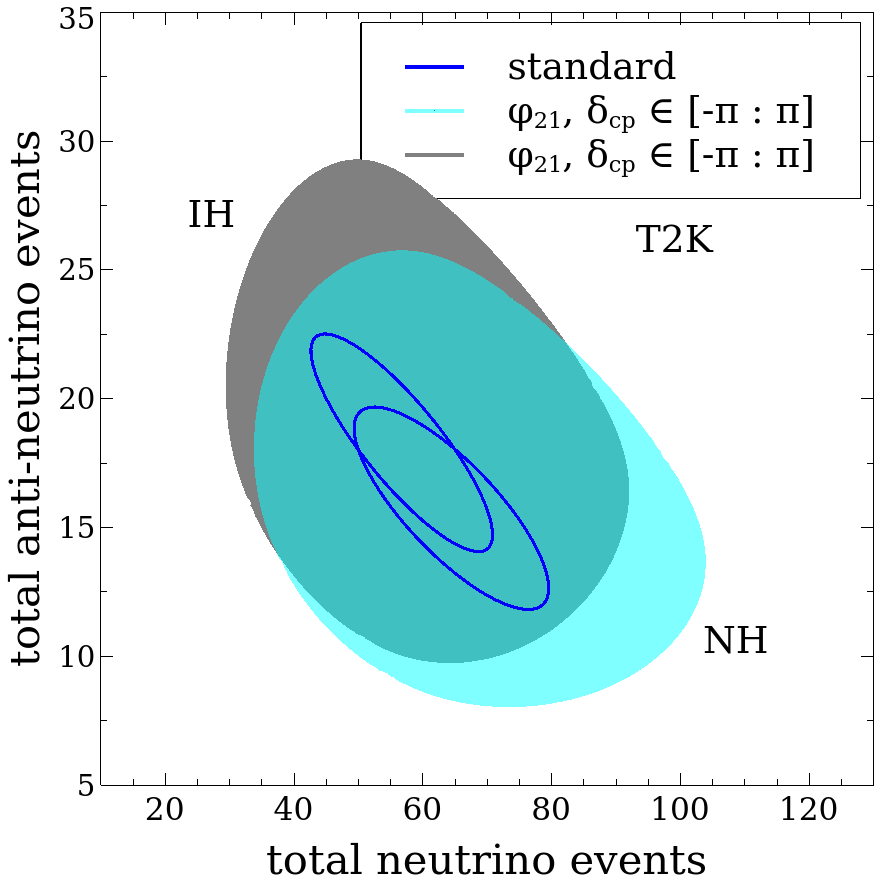}
\caption{\footnotesize{Energy dependent bi-event plots For DUNE, NO$\nu$A and T2k for their whole energy ranges i.e from 0.5 to 10 GeV, 0.4 to 4 GeV and 0.4 to 1.2 GeV respectively. All other variations are same as the previous plot ( figure \ref{event1}).}}
\label{event2}
\end{figure}

\item NO$\nu$A and T2K show another interesting feature. In NO$\nu$A, almost 50$\%$ of the standard ellipse in a particular hierarchy is a part of the NU induced band of ellipses in the opposite hierarchy. So in that region, any co-ordinate in the standard blue ellipse is not only a part of the standard NH ellipse, but also a part of both the gray and cyan bands.  In T2K, similar behavior  can be seen for a larger region of the parameter space.

 \end{itemize}
 
 From the above analysis, it is clear that in presence of NU, all the three experiments are incapable of discriminating between the two mass hierarchies. DUNE can tell us about the mass hierarchy when operating at its peak energy as seen from figure~\ref{prob} and \ref{event1}. Still there is tension between the two hierarchies in the presence of NU. In the next section, we discuss these issues in terms of sensitivity plots.

\section{Sensitivity Studies}
\subsection{Statistical Details and $\chi^2$ Analysis}
The results presented in this section are based on $\chi^2$ analysis where we have calculated $\Delta\chi^2$ by comparing the predicted spectra for the alternate hypothesis. For an assumed normal hierarchy as the true hierarchy, $\Delta\chi^2_{\rm MH}$ is defined as $\chi^2_{\rm NH} -\chi^2_{\rm IH}$. Similarly for an assumed true inverted hierarchy, $\Delta\chi^2_{\rm MH} =\chi^2_{\rm IH} -\chi^2_{\rm NH}$. Now, in terms of event rates, we can define it as: 
\begin{equation}
\chi^2 = \sum_{i=1}^{\rm bins} \sum_{j=1}^{2}\frac{[N^{i,j}_{\rm true} - N^{i,j}_{\rm test}]^2}{N^{i,j}_{\rm true}},
\end{equation}
 where $N^{i,j}_{\rm true}$ and $N^{i,j}_{\rm test}$ are the event rates that correspond to data and fit in the $i^{th}$ bin. $j = 1$ is for neutrinos and $j = 2$ for anti-neutrinos. The number of bins are different for each experiment i.e. for DUNE there are 39 bins each of width 250 MeV in the energy range 0.5 to 10 GeV, for NO$\nu$A there are 28 bins of width 125 MeV in the energy range 0.5 to 4 GeV and for T2K, we have 20 bins of width 40 MeV in the range 0.4 to 1.2 GeV. 

 In the $\chi^2$ calculation for the standard 3$\nu$ case, we have marginalised over the whole range of $\delta_{cp}$ from $-\pi$ to $\pi$ in the `fit'. To measure the hierarchy sensitivity, we fix our `data' in a particular hierarchy and test the opposite hierarchy in the `fit'. We have also marginalised over $\Delta m^2_{31}$ in the `fit' in its allowed 3$\sigma$ ranges i.e. for an assumed NH as the true hierarchy, we vary $\Delta m^2_{31}$ in the `fit' assuming IH. Then we calculate the minimized $\chi^2$ (i.e. $\chi^2_{\rm min}$) for each true $\delta_{cp}$ assuming the best fit values of the oscillation parameters as the true values. In the presence of NU, in addition to $\dcp$ and $\Delta m^2_{31}$, we have marginalised over all the non unitarity parameters in the `fit' in their allowed ranges assuming the central values as the true values. For a particular true value of $\delta_{cp}$ we show the maximum and the minimum of $\chi^2_{\rm min}$ which is obtained corresponding to a variation of the new phase $\phi_{21}$ in the `data' from  $-\pi$ to $\pi$.

\subsection{Mass Hierarchy Sensitivity}

Here we present our results for the mass hierarchy sensitivity of the three experiments in the presence of NU. We compare our results with the standard three flavor case. We also show the combined hierarchy sensitivity of the T2K and NO$\nu$A experiments in the presence of NU. In the plots, the blue line corresponds to the standard hierarchy sensitivity of these experiments while the gray band shows the effect of NU. The green line corresponds to the special case where $\phi_{21}$ is zero in both `data' and `fit' i.e. it shows the effect of the three absolute NU parameters $\alpha_{11}$, $\alpha_{22}$ and $|\alpha_{21}|$. We have shown the sensitivity plots for both the hierarchies. We have combined both $\nu_{e}$ appearance and $\nu_{\mu}$ disappearance channels in both the $\nu$ and $\bar{\nu}$ modes to utilize the full potential of each experiment towards mass hierarchy measurements. We can make the following observations from these plots:
\begin{figure}[H] 
\centering
\includegraphics[width=0.69\textwidth]{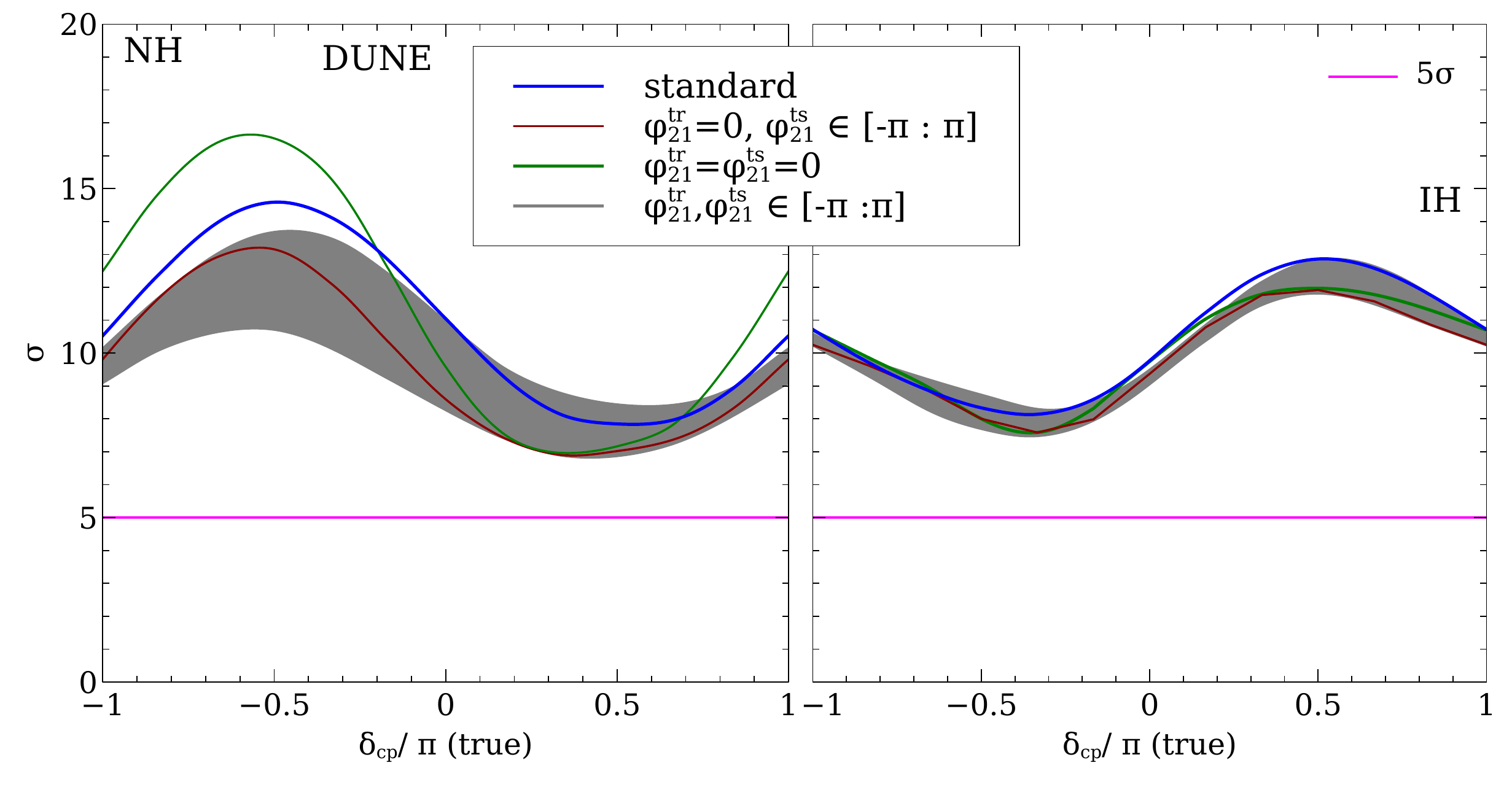}
\includegraphics[width=0.69\textwidth]{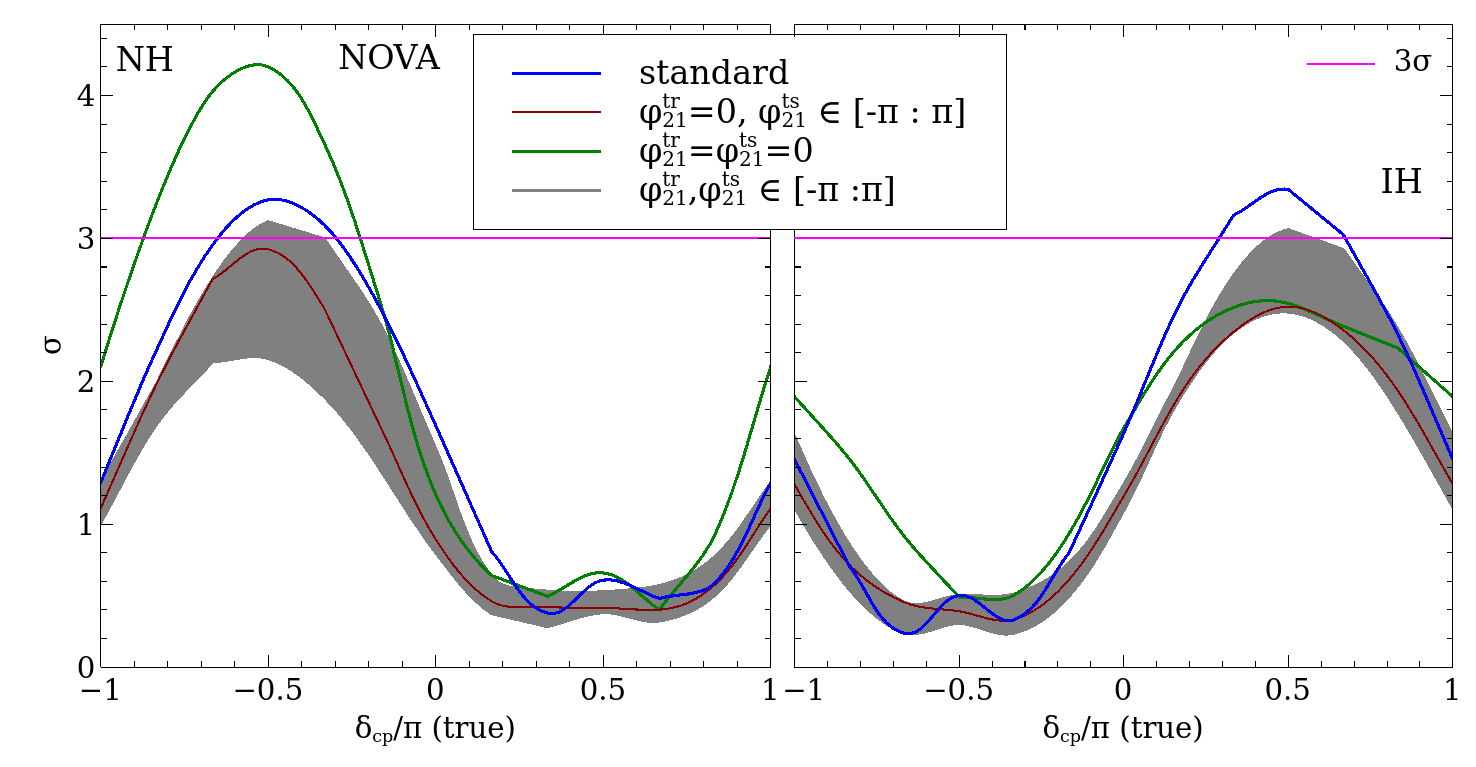}
\includegraphics[width=0.69\textwidth]{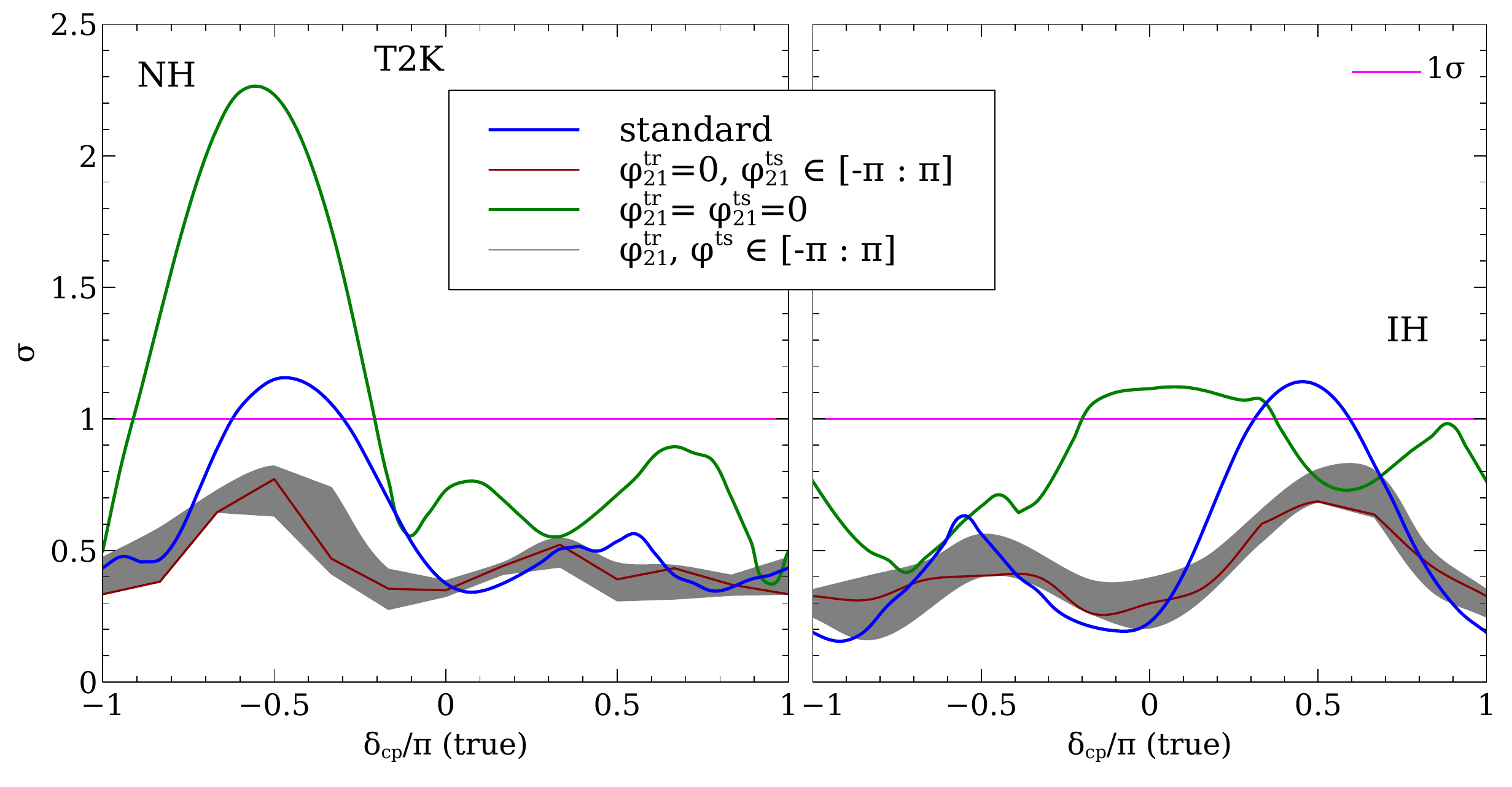}
\caption{\footnotesize{ Mass hierarchy sensitivity plots for DUNE (5+5), No$\nu$A (3+3) and T2K (3+3) for both the hierarchies. The blue line represents the standard mass hierarchy sensitivity. The gray band corresponds to the variation of true $\dcp$ and $\phi_{21}$ (both $\phi^{tr}_{21}$ $(\rm true)$ and $\phi^{ts}_{21}$ $(\rm test)$) from $[-\pi, \pi]$ . The dark red line represent the case when $\phi^{tr}_{21} = 0$ but $\phi^{ts}_{21}$ is varied from $[-\pi, \pi]$. The green line show the effect of the non zero absolute parameters for $\phi^{tr}_{21}$ = $\phi^{ts}_{21}$ = $0$. For all the cases with NU, we assume the central values of the NU parameters as the true values.}}
\label{dune}
\end{figure}

\begin{itemize}
\item It is seen in the bi-probability and bi-event plots for DUNE that the ellipses corresponding to both the hierarchies are well separated in the standard three flavor case.  figure \ref{dune} shows that for an assumed NH as the true hierarchy, DUNE can exclude the wrong hierarchy (i.e. IH in this case) at more than 5$\sigma$ C.L. for all the true values of $\delta_{cp}$. But in the presence of NU, for the true NH case, the mass hierarchy sensitivity decreases in the lower half plane (LHP, from $-\pi$ to 0) compared to the standard scenario. In the upper half plane (UHP, from 0 to $\pi$), the sensitivity with NU increases compared to the standard case for some fraction of true $\delta_{cp}$, especially near $\delta_{cp} = \pi$.

The effect of marginalizing over a large parameter space brings the $\chi^2$ down and hence in the presence of NU, the mass hierarchy sensitivity decreases. In the case of NO$\nu$A, the hierarchy sensitivity in the standard scenario is already less than 3$\sigma$ except near $\dcp =-\pi/2$, while that of T2K is less than 2$\sigma$. In the presence of NU, this sensitivity further decreases especially in the LHP for an assumed true NH. IN the UHP, the hierarchy sensitivity in the presence of NU increases for some true combinations of $\dcp$ and $\phi_{21}$, but the increase is not so significant.  

\begin{figure}[H] 
\centering
\includegraphics[width=0.7\textwidth]{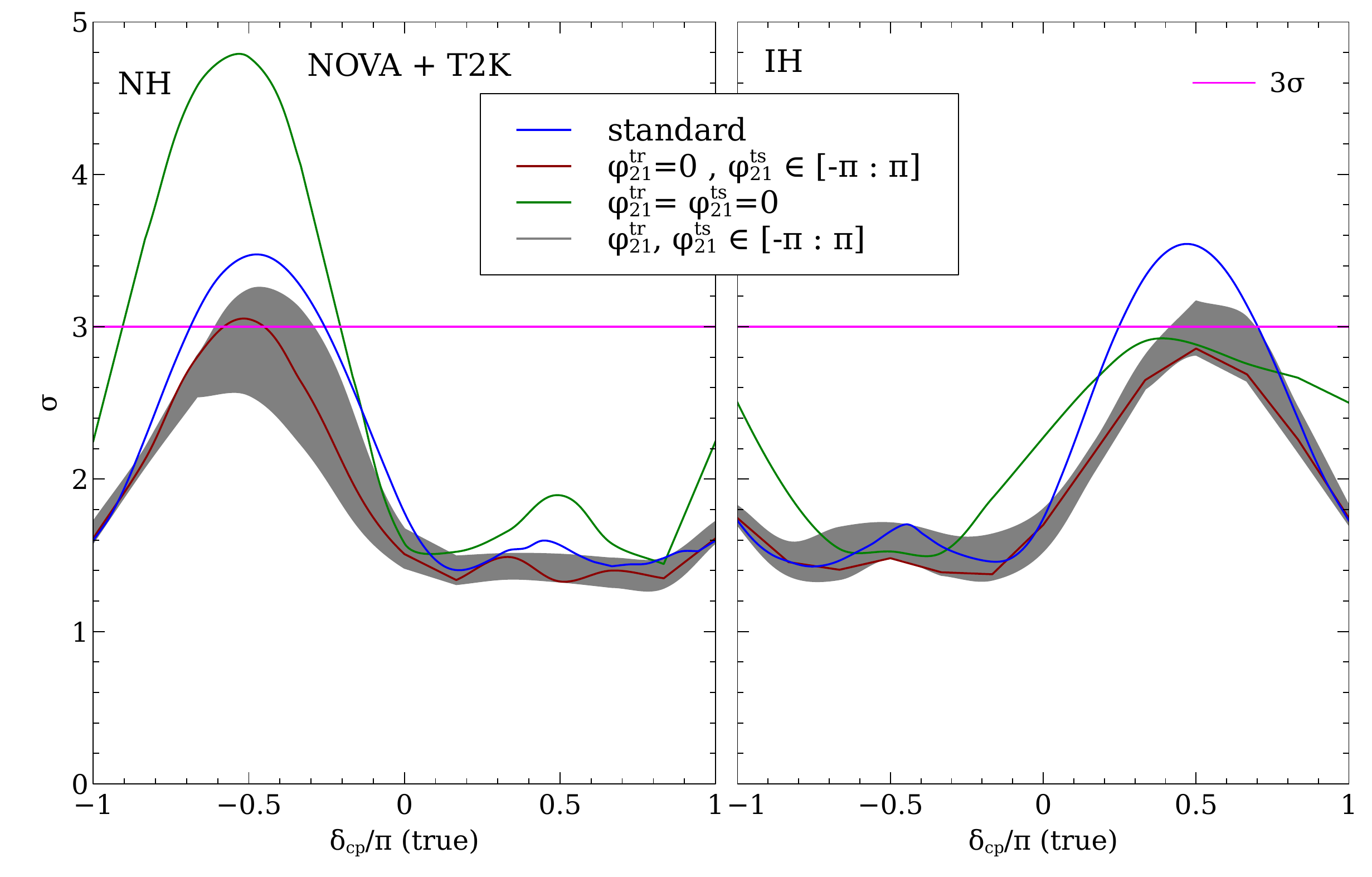}
\caption{\footnotesize{Mass hierarchy sensitivity plots for the combination of NO$\nu$A (3+3) and T2K (3+3) in both the hierarchies. The blue line represents the standard mass hierarchy sensitivity. The gray band corresponds to the true variation of $\dcp$ and $\phi_{21}$ from $[-\pi, \pi]$. }}
\label{novat2k}
\end{figure}

\item The dark red plot, representing the special case of true $\phi_{21} = 0$, lies within the gray band for all the three experiments as expected. In DUNE, the green plot, showing the sensitivity when both true and test $\phi_{21} = 0$ (if there is no new physics phase), shows higher sensitivity for $\dcp \in$ $[-\pi, -\pi/6]$ and $\dcp > 1.45\pi$ than the standard case for an assumed true NH. But in between $\dcp \in$ $[-\pi/6, 1.45\pi]$ the green plot dips below the standard 3$\nu$ scenario. In the case of NO$\nu$A, the green plot drops compared to the standard 3$\nu$ sensitivity only for a small fraction of $\dcp$ around $0$ and $1.5\pi$ for an assumed true NH. The sensitivity shoots up to 3$\sigma$ for more than 70$\%$ of true $\dcp$ in the LHP. In the case of T2K, the green line is always higher than the standard sensitivity for assumed true NH. For more than 70\% of true $\dcp$ in the LHP, sensitivity is higher than 1$\sigma$.

\item For an assumed true IH, DUNE can exclude NH for all values of true $\delta_{cp}$ at more than 5$\sigma$ C.L.. Even in the presence of NU, DUNE can resolve the neutrino mass hierarchy at more than 5$\sigma$ C.L. irrespective of the true hierarchy. But in the case of NO$\nu$A and T2K, the sensitivity decreases with NU and T2K is the most affected.

\item figure \ref{novat2k} shows the combined hierarchy sensitivity of T2K and NO$\nu$A. The sensitivity increases slightly compared to their individual sensitivities. In the presence of NU, some fraction of $\dcp$ around $-\pi$ ($\pi$) has a sensitivity more than 3$\sigma$ in the NH (IH) case. We have not combined DUNE data with T2K and NO$\nu$A as its individual sensitivity is more than 5$\sigma$.
\end{itemize}


\subsection{Mass Hierarchy Discovery Reach}

In this section, we show the mass hierarchy discovery reach of these experiments in the presence of NU. As the DUNE experiments can rule out the wrong hierarchy with more than 5$\sigma$ C.L. even in the presence of NU, here we present the discovery potential of T2K and NO$\nu$A and their combinations only. We have generated these results assuming the maximum deviation from unitarity, i.e. the true values of the NU parameters are fixed at their boundary values. The contours are shown in $\dcp-\phi_{21}$ (true) parameter space and  are drawn at 1 d.o.f.. The regions bounded by the contours are the allowed regions in this parameter space. 

\begin{figure}[H] 
\centering
\includegraphics[width=0.45\textwidth]{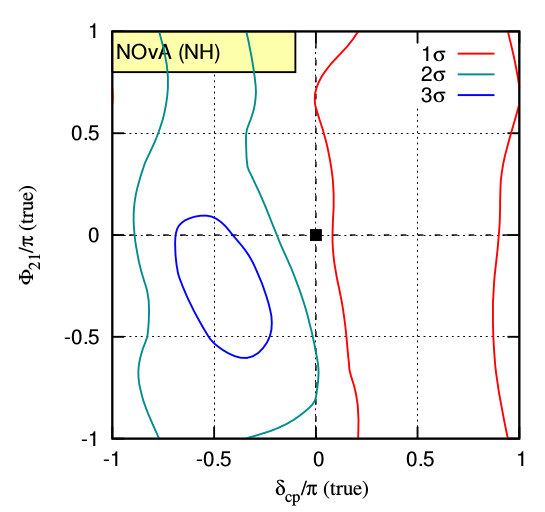}
\includegraphics[width=0.44\textwidth]{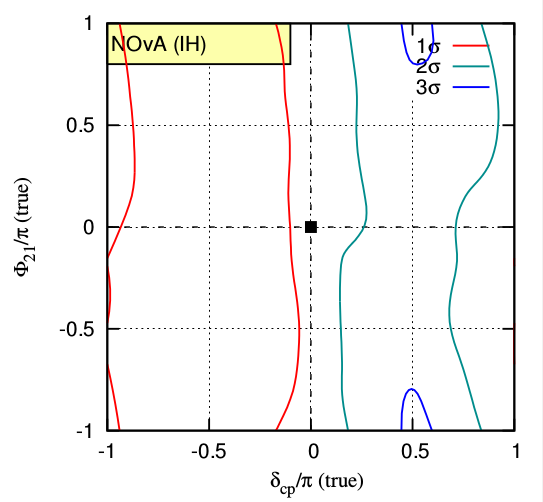}
\includegraphics[width=0.45\textwidth]{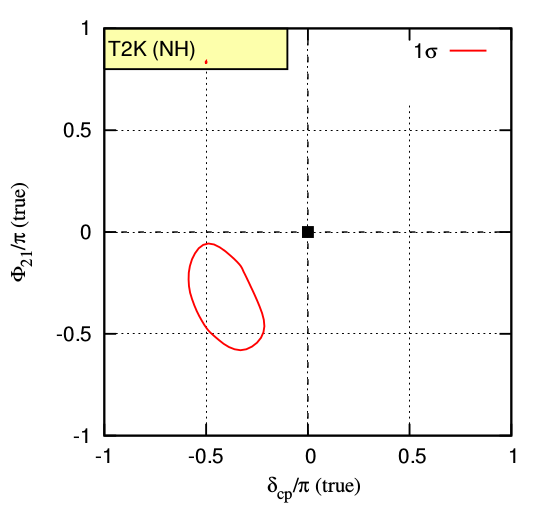}
\includegraphics[width=0.45\textwidth]{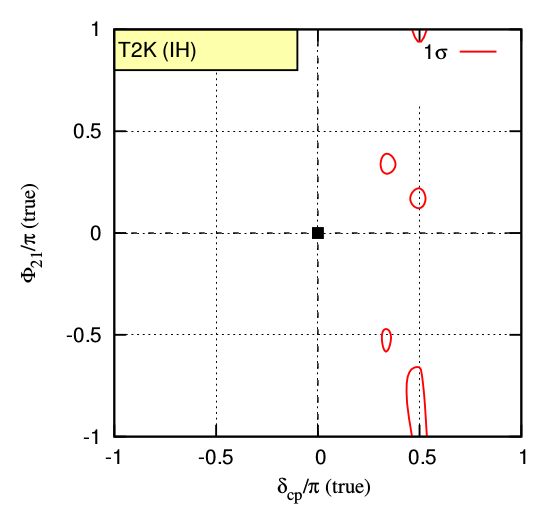}
\caption{\footnotesize{ Mass hierarchy discovery potential of NO$\nu$A (3+3) and T2K (3+3) for both the hierarchies. Here we assume the boundary values of the NU parameters as the true values i.e.$\alpha_{11} = 0.9945$, $\alpha_{22} = 0.9995 $ and $|\alpha_{21}| = 0.0257$.  }}
\label{cont1}
\end{figure}

In figure \ref{cont1}, we show the MH discovery reach of NO$\nu$A and T2K for both NH and IH. For each true $\Delta m^2_{31}$ in NH (IH), we vary test $\Delta m^2_{31}$ in IH (NH). We observe that NO$\nu$A can probe NH at 3$\sigma$ C.L. for some true combinations of $\dcp$ and $\phi_{21}$. The region outside the blue contours is the excluded region where 3$\sigma$ discovery of MH is not possible. The 3$\sigma$ allowed region shrinks for the case where assumed true hierarchy is inverted. In the case of T2K, only a 1$\sigma$ discovery is possible with NU for both the hierarchies.
\begin{figure}[H] 
\centering
\includegraphics[width=0.45\textwidth]{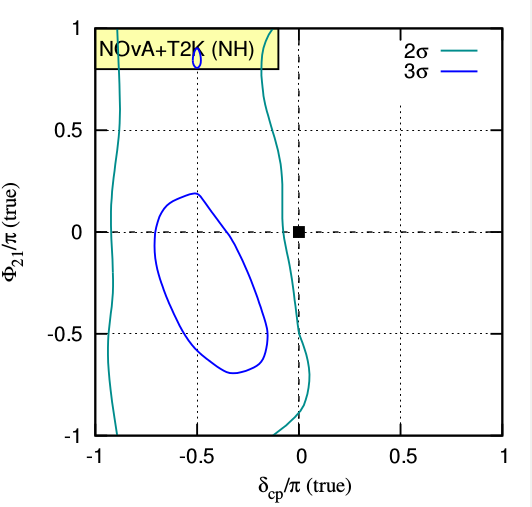}
\includegraphics[width=0.45\textwidth]{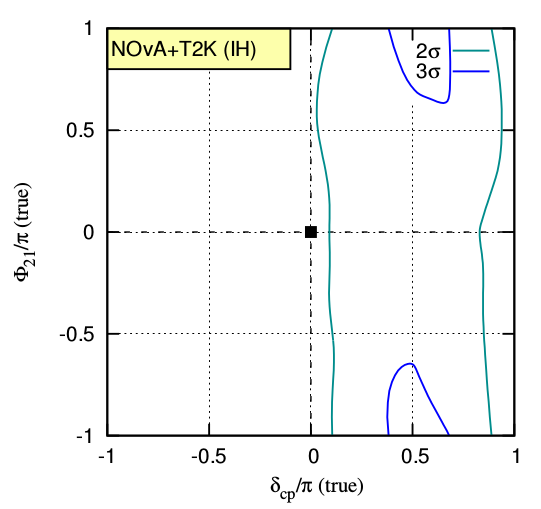}
\caption{\footnotesize{ Mass hierarchy discovery potential of the combined experiments : We combine No$\nu$A (3+3) and T2K (3+3) data in both the hierarchies. Here also, we assume the boundary values of the NU parameters as the true values i.e.$\alpha_{11} = 0.9945$, $\alpha_{22} = 0.9995 $ and $|\alpha_{21}| = 0.0257$. }}
\label{cont2}
\end{figure}
In figure \ref{cont2}, we have presented our results for the combined case. Here we observe that adding T2K data with NO$\nu$A can slightly improve the discovery potential of NO$\nu$A. The size of the blue contours increases slightly for both the hierarchies compared to NO$\nu$A, which in turn means that for more true values of $\dcp$ and $\phi_{21}$, the combined setup can discover MH at 3$\sigma$ C.L..

\begin{figure}[] 
\centering
\includegraphics[width=0.45\textwidth]{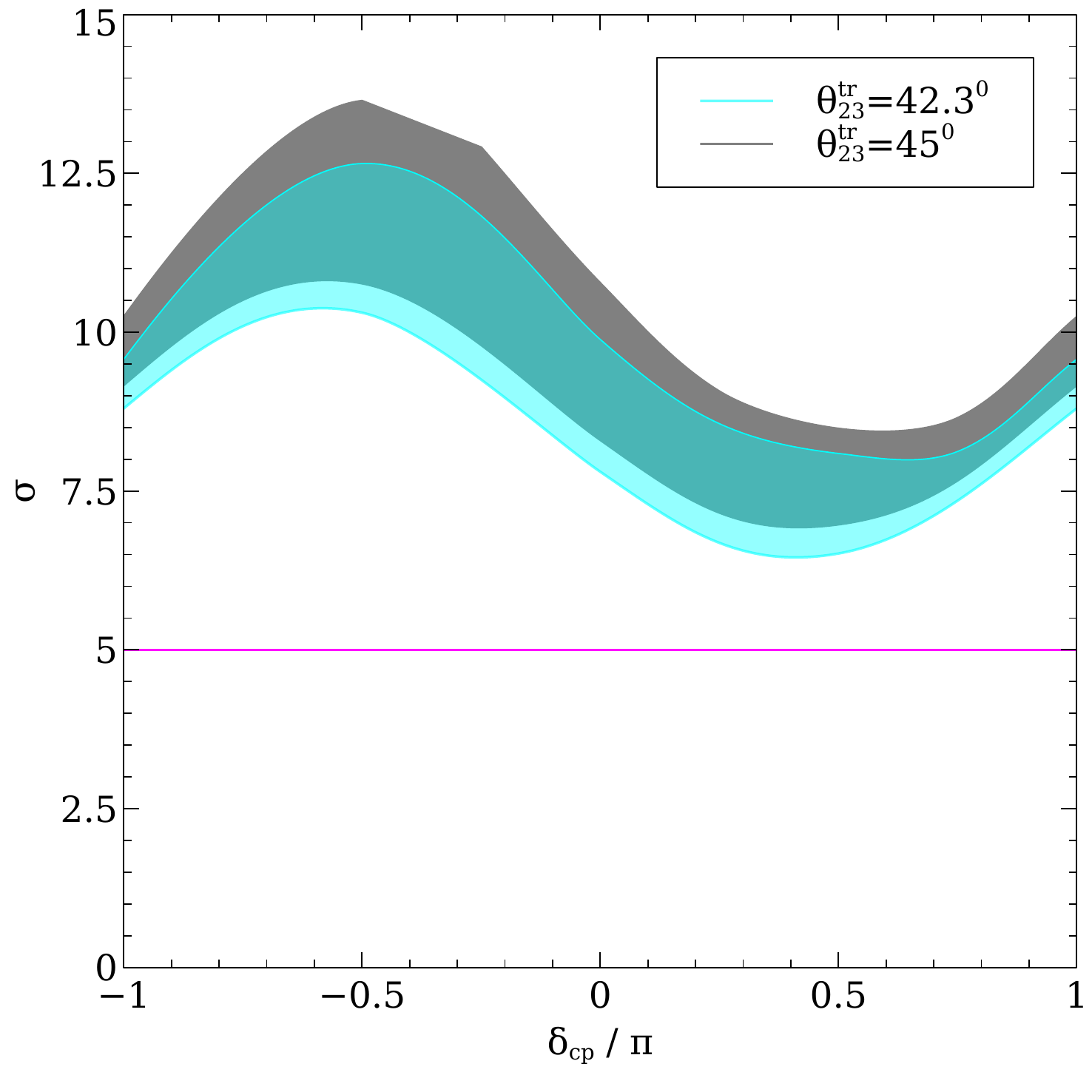}\\
\includegraphics[width=0.45\textwidth]{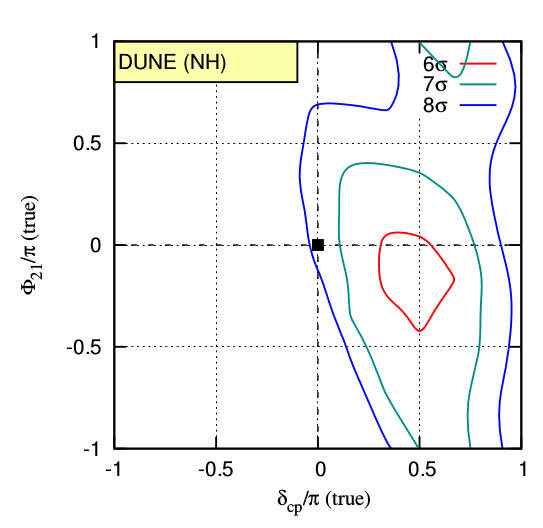}
\includegraphics[width=0.45\textwidth]{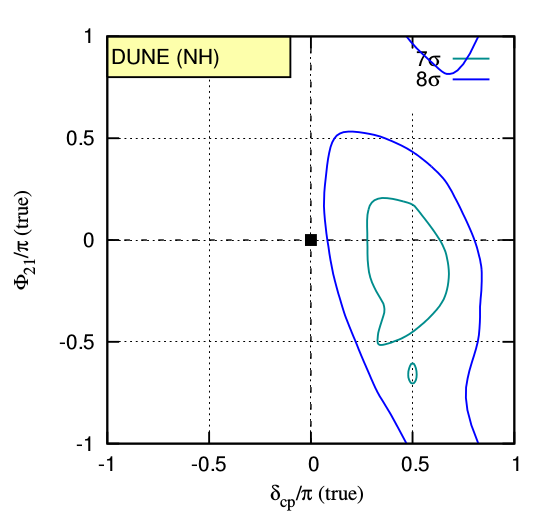}
\caption{\footnotesize{ Here, upper panel shows the mass hierarchy sensitivity of DUNE for true $\tz=45^0 $ and $\tz=42.3^0$ marginalising $\tz$ in the allowed 3$\sigma$ range in fit. Lower panel shows the contour plots for $\tz=38.3^0$ (left) and $\tz=42.3^0$ (right) in the true $\dcp-\phi_{21}$ parameter space.}}
\label{new}
\end{figure}

 In fig.\ref{new}, we have shown the effect of marginalisation on $\theta_{23}$ for maximal and non maximal true $\tz$ and have observed that the physics conclusions drawn in this work are not going to change with non maximal true $\tz$. Also from the contour plots in the lower panel, it is confirmed that the NU effect seen in probability level at DUNE ( APPENDIX 1) can be lifted by combining appearance and disappearance channels both in neutrino and anti-neutrino mode. 
\section{Conclusions}
In this work, we have attempted to analyze the mass hierarchy sensitivity of the long-baseline experiments  T2K, NO$\nu$A and DUNE
in the presence of non-unitarity. Below we summarize the salient conclusions of this work:

\begin{itemize}

\item The presence of non-unitarity leads to a strong degeneracy between the standard 3$\nu$ case and the NU induced case in estimating the neutrino mass hierarchy for all three superbeam experiments at the level of oscillation probabilities.
An analysis of the bi-probability plots of the three experiments shows that only DUNE can discriminate between the two hierarchies at its peak energy even in the presence of NU, while T2K and NO$\nu$A show significant overlaps between the NH and IH ellipses. 
Also, for certain values of $\delta_{cp}$ and the NU phase in all three experiments, it is not possible to specify whether the value arises from the standard case or the NU induced case. 

\item  From the bi-event plots for the three experiments, we observe that if analyzed at the peak energies of the respective experiments, DUNE can distinguish between the hierarchies even in the presence of NU, while NO$\nu$A and T2K are unable to do so because of their shorter baselines and less matter effects. If an integration over the energy ranges of the experiments is taken into account, then DUNE also suffers from a small overlap between the hierarchies with NU. Further, for NO$\nu$A and T2K in the presence of NU, there is a degeneracy in the same hierarchies between the standard and the NU induced hierarchy measurements, as well as a degeneracy between NH (IH) in the 3$\nu$ scenario and IH (NH) in the NU induced scenario. Also, with NU any of these experiments may misinterpret a non unitary event as a standard 3$\nu$ event.
Thus at the event level, all the three experiments are incapable of discriminating between the mass hierarchies with NU, except for DUNE at its peak energy.

\item The results for the sensitivity to the mass hierarchy show that with NH as the true hierarchy, DUNE can exclude the wrong hierarchy at more than 5$\sigma$ C.L. for all true values of $\delta_{cp}$. But in the presence of NU, for the true NH case, the mass hierarchy sensitivity decreases in the LHP ($-\pi$ to 0) compared to the standard scenario. In the UHP (0 to $\pi$), the sensitivity with NU increases compared to the standard case for some combination of true $\delta_{cp}$ and $\phi_{21}$, especially near $\delta_{cp} = \pi /2$. In the case of NO$\nu$A, the hierarchy sensitivity in the standard scenario is already less than 3$\sigma$ except near $\dcp =-\pi/2$, while that of T2K is less than 2$\sigma$. In the presence of NU, this sensitivity further decreases especially in the LHP for an assumed true NH. IN the UHP, the hierarchy sensitivity in the presence of NU increases for some true combinations of $\dcp$ and $\phi_{21}$, but the increase is not significant.

\item For true IH, DUNE can exclude NH for all values of true $\delta_{cp}$ at more than 5$\sigma$ C.L.. Even in the presence of NU, DUNE can resolve the neutrino mass hierarchy at more than 5$\sigma$ C.L. irrespective of the true hierarchy. But in the case of NO$\nu$A and T2K, the sensitivity decreases with NU and T2K is the most affected.

\item The combined hierarchy sensitivity of T2K and NO$\nu$A increases slightly compared to their individual sensitivities. In the presence of NU, some fraction of $\dcp$ around $-\pi/2$ ($\pi/2$) has a sensitivity more than 3$\sigma$ in the NH (IH) case. 

\item Finally, the mass hierarchy discovery reach of NO$\nu$A and T2K is studied and it is observed that NO$\nu$A can probe NH at 3$\sigma$ C.L. for some true combinations of $\dcp$ and $\phi_{21}$. The 3$\sigma$ allowed region shrinks for the case true IH. In the case of T2K, only 1$\sigma$ discovery is possible with NU for both the hierarchies. Adding T2K data with NO$\nu$A can slightly improve the discovery potential of NO$\nu$A. The combined setup can discover MH at 3$\sigma$ C.L.. for a greater range of true values of $\dcp$ and $\phi_{21}$. We have not studied the discovery reach for DUNE since it has already been observed that it can rule out the wrong hierarchy at more than 5$\sigma$ C.L. even with NU.

\item We have carefully checked our results (figure \ref{dune} and \ref{cont1}) for non maximal $\theta_{23}$ values (with the present best fit value and and a benchmark value of $\theta_{23}=38.3^{0}$ as the true value), marginalizing over the whole allowed range of $\theta_{23}$ in the fit for DUNE (for true NH). In figure \ref{new}, we depict the DUNE hierarchy sensitivity showing the comparison between maximal and non-maximal $\theta_{23}$ (best fit value $\theta_{23}=42.3^{0}$). In the lower panel, we have shown the contour plots for true $\theta_{23}=42.3^{0}$ (left) and true $\theta_{23}=38.3^{0}$ (right). We have chosen $\theta_{23}=38.3^{0}$ just to show the maximal possible correlation between $\phi_{21}$ and $\delta_{cp}$ The contour plot with  $\theta_{23}=42.3^{0}$ is consistent with the sensitivity plot shown in the upper panel.
We observe from the upper panel that the differences between the sensitivity for maximal and non-maximal $\theta_{23}$ are very small and in all cases, the capability of DUNE to exclude the wrong IH is much more than 5$\sigma$ for all true values of $\delta_{cp}$. This is also the reason why the whole $\delta_{cp} -\phi_{21}$ parameter space is excluded for DUNE for assumed true NH/IH at more than 5$\sigma$ C.L., as confirmed by the contour plots in the lower panel. The reason for this is as follows: the probability plots shown in Appendix 1 are only in the $\nu$ mode. But when we combine both $\nu$ and $\bar{\nu}$ in appearance and disappearance modes to calculate the sensitivity, NU hampers the sensitivity at DUNE at a higher confidence level only. And also, as pointed out in Section III, the physics conclusions drawn here remain unchanged even if we consider non maximal $\theta_{23}$ in `data' and then marginalize it in `fit' in the allowed 3$\sigma$ range.

\end{itemize}

We conclude that the presence of non-unitarity in the neutrino mass matrix can significantly affect the potential of the experiments NO$\nu$A and T2K to resolve the neutrino mass hierarchy. The experiment DUNE is less affected due to its longer baseline and consequent large matter effects, which results in a resolution of the hierarchy degeneracy for DUNE even in the presence of non-unitarity at its peak operating energy. It is worthwhile to analyze other long baseline experiments to understand more thoroughly the effect of non-unitarity on their capability for determining the mass hierarchy. 

\section{Appendix 1}
\begin{figure}[h] 
\centering
\includegraphics[width=0.9\textwidth]{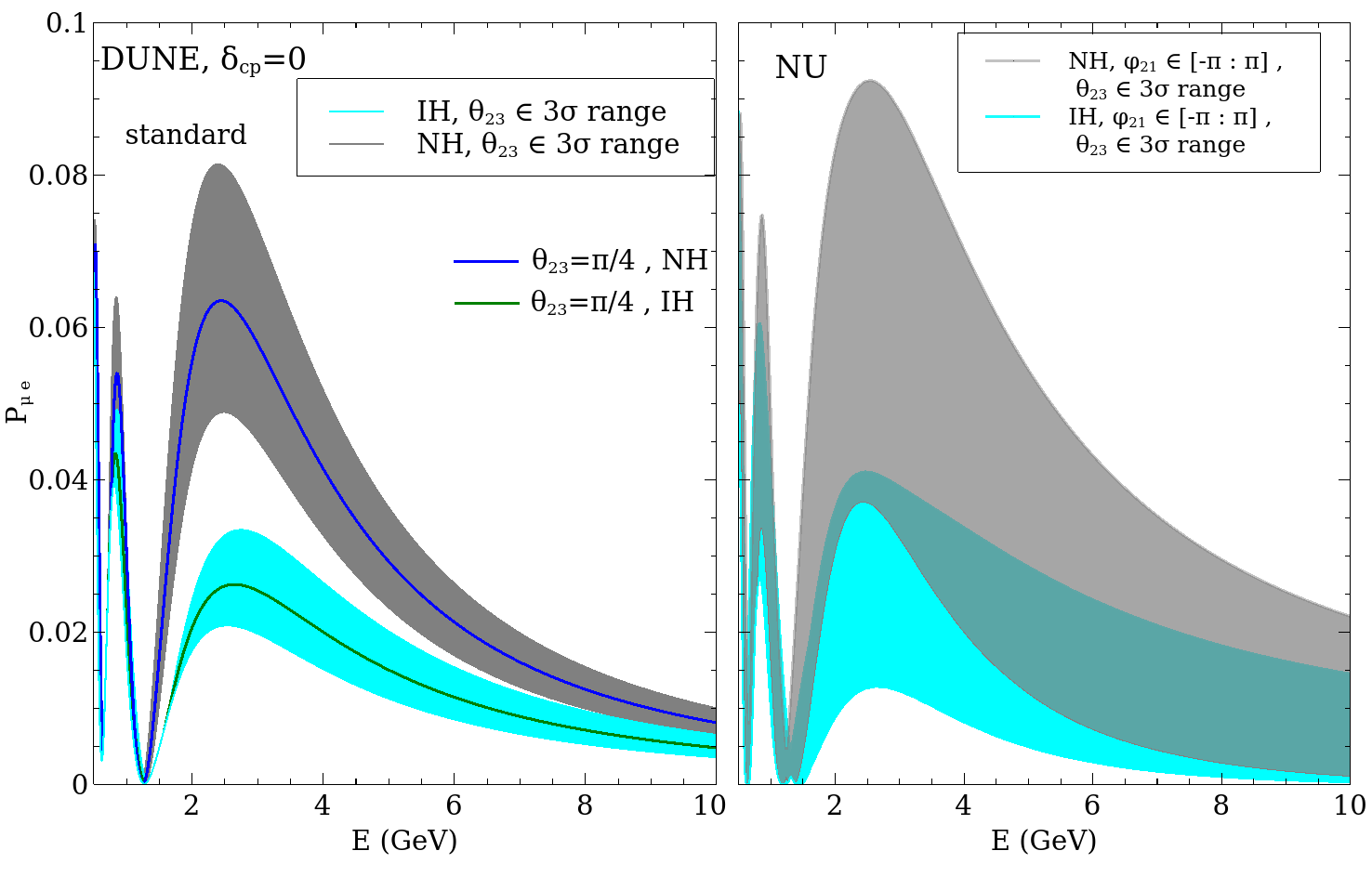}
\caption{\footnotesize{ $\rm P_{\mu e}$ vs Energy plots for DUNE to show the effect of $\tz$ variation. In the left (right) panel, we show the variation in standard ( with NU) case. We consider the boundary values of the NU parameters here. The blue (green) line represent the $\rm P_{\mu e}$ vs E for $\tz=45^0$ in NH (IH). The gray (cyan) band show the variation of $\tz$ in 3$\sigma$ allowed range in NH (IH) mode. }}
\label{th23}
\end{figure}
In this study we have assumed the maximal value of the atmospheric neutrino mixing angle $\theta_{23}$. However it can be shown at the probability level that there is a significant effect of varying $\theta_{23}$ on the oscillation probability, which has the potential of 
affecting the results for the hierarchy sensitivity. To demonstrate this we present in figure \ref{th23} the probability $P_{\mu e}$ as a function of the neutrino energy for DUNE, incorporating a variation in $\theta_{23}$ depicted by the grey (cyan) band for true NH (IH) in the figure. Here we have compared the standard case with the NU case for $\dcp=0$. In the left panel, we vary $\theta_{23}$ over its current 3$\sigma$ range for both the hierarchies and see that the probabilities for NH and IH  are still well separated. But in the presence of NU (right panel), if we vary over $\theta_{23}$, there is a large overlapping region between the probabilities for NH and IH. This indicates that a more rigorous procedure should be followed to take into account the current uncertainty in $\theta_{23}$ while performing this analysis.

\begin{acknowledgments}
We thank Prof. Raj Gandhi for his useful comments and suggestions in the manuscript. We acknowledge the use of HRI cluster facility to carry out the computations. DD acknowledges the support from the DAE Neutrino project at HRI. PG acknowledges local support for research at LNMIIT, Jaipur. SR thanks Mehedi Masud and Sandeep K. Sehrawat for useful discussions.
\end{acknowledgments}

\end{document}